# Transport and magnetic properties of Co-doped $BaFe_2As_2$ epitaxial thin films

Shyam Mohan[1], Toshihiro Taen[1], Hidenori Yagyuda[1], Yasuyuki Nakajima[1,2], Tsuyoshi Tamegai[1,2], Takayoshi Katase[3], Hidenori Hiramatsu[4] and Hideo Hosono[3,4]

[1]Department of Applied Physics, The University of Tokyo, 7-3-1 Hongo, Bunkyo-ku, Tokyo 113-8656, Japan
[2]JST, Transformative Research-Project on Iron Pnictides (TRIP), 7-3-1 Hongo, Bunkyo-ku, Tokyo 113-8656, Japan
[3]Materials and Structures Laboratory, Mailbox R3-1, Tokyo Institute of Technology, 4259 Nagatsuta, Midori, Yokohama 226-8503, Japan
[4]Frontier Research Center, S2-6F East, Mailbox S2-13, Tokyo Institute of Technology, 4259Nagatsuta, Midori, Yokohama 226-8503, Japan

**Abstract.** We report resistivity, Hall coefficient, current-voltage characteristics, and magneto-optical imaging measurements of epitaxial Co-doped $BaFe_2As_2$ thin films deposited on MgO(001) substrate. The Hall resistivity of the films has a substantial contribution arising from anomalous Hall effect of ferromagnetic components. The critical current density ($J_c$) of the films is $\sim 2$ $MA/cm^2$ at low temperatures. Differential magneto-optical images of the remanent state give similar $J_c$ values and also exhibit presence of extended defects in the film.

The discovery of iron-based superconductors [1] has led to an intensive research activity with an aim to understand their fundamental properties as well as to pursue potential applications. Thin films of various iron-based superconductors have attracted attention for potential electronic device applications and studying the intrinsic properties. Thin films of F-doped $Ln$FeAsO ($Ln$ = La,Nd) [2,3,4,5,6], Co-doped $AE$Fe$_2$As$_2$ ($AE$ = Ba, Sr) [7,8,9,10,11,12], and FeSe$_{1-x}$Te$_x$ [13,14,15] have been successfully prepared by various groups. The difficulty in controlling the stochiometry coupled with the high vapour pressure of As at high temperatures caused considerable difficulties in the fabrication of LaFeAs(O,F) thin films: the initial films were not superconducting [2]. Subsequently, it was found that post-annealing of the LaFeAs(O,F) films leads to appearance of superconductivity [3]. Very recently, successful fabrication of NdFeAs(O,F) thin films showing superconductivity at 48 K without post-annealing has been reported [6]. On the other hand, thin films of Co-doped $BaFe_2As_2$ have become a prime candidate for device fabrication [16] due to the ease of growing films with modest transition temperatures ($T_c$), high crystallinity and high chemical stability in ambient atmosphere [8]. While Co-doped $BaFe_2As_2$ films deposited on SrTiO$_3$ exhibited high critical current density ($J_c$) > 1 $MA/cm^2$, the conducting paths in SrTiO$_3$ was a concern [11]. Subsequently, films deposited on insulating (La,Sr)(Al,Ta)O$_3$ using an intermediate layer of SrTiO$_3$ or BaTiO$_3$ were shown to be truly epitaxial and possessing high $J_c$ $\sim 4$ $MA/cm^2$ [12]. However, careful optimization of the deposition conditions also lead to such high values of $J_c$ for films deposited on (La,Sr)(Al,Ta)O$_3$ without the need for buffer layers [17]. The $T_c$ of films deposited by various groups are close to that of Co-doped $BaFe_2As_2$ single crystals. In particular, transition temperature of the films were found to increase with increasing $c/a$ ratio

[10], with $T_c$ reaching about 24 K for films deposited on (La,Sr)(Al,Ta)O$_3$ and SrTiO$_3$ and having $c/a$ ~3.3.

Here we present the characterization of superconducting properties of epitaxial thin films of Co-doped BaFe$_2$As$_2$ deposited on MgO substrates. The results of our investigations using resistivity, Hall coefficient, current-voltage (*I-V*) characteristics, and magneto-optical imaging measurements are discussed.

8% Co-doped BaFe$_2$As$_2$ films with thickness 220 nm were deposited on MgO(001) single-crystal substrates by laser ablation [17]. The phase purity and crystallinity were quite similar to those of samples reported in Ref. 17. The obtained films were patterned in a 6.7 μm wide four-terminal and 0.1 mm wide six-terminal shape for *I-V* measurements, and for resistivity and Hall measurements, respectively by photolithography and Ar ion milling.

For transport measurements, terminal contacts were made with gold wires using conductive silver paint. Resistivity measurements were performed in the standard four-probe method with the current applied along the *ab* plane of the film. The distance between voltage contacts was 650 μm. Figure 1(a) shows the temperature (*T*) dependence of the in-plane resistivity ($\rho_{xx}$) in the film at zero-field. The room temperature resistivity of the film is $\rho_{xx}$ (300 K) = 175 μΩcm; this is much lower than the corresponding value (~300 μΩcm) reported in single crystals [18]. With decreasing temperature, the resistivity decreases monotonically followed by an abrupt drop starting at $T_{c,onset}$ ~21 K and reaching zero resistivity at $T_{c,zero}$ ~19 K. The residual resistivity ratio $\rho_{xx}$ (300 K)/ $\rho_{xx}$ ($T_{c,onset}$) is 1.94 which compares well with the value in single crystals [18]. The upper inset in figure 1 shows the variation of the magnetization with temperature, measured using a commercial SQUID magnetometer (Quantum Design MPMS-XL5), both in the zero-field-cooled (ZFC) and field-cooled (FC) modes with a field *H* = 5 Oe applied along the *c*-axis of the film. A gradual transition to the superconducting state starting from ~18 K is observed. The variation of $T_c$ with magnetic field is shown in the lower inset of figure 1(a) for *H* = 0, 10, 20, 30, 40, and 50 kOe applied along the *c*-axis of the film. With increasing field, the resistive transition shifts to lower temperatures accompanied by a slight increase in the transition width. Figure 1(b) (filled symbols) shows the variation of the upper critical field $H_{c2}$ along the *c*- and *ab*-directions with reduced temperature $t = T/T_{c,onset}$. The values of $H_{c2}$ were defined as the field at the midpoint of the resistive transition. The slopes of $H_{c2}$ along the *c*- and *ab*-directions at $T_c$ are -28.2 kOe/K and -51.3 kOe/K, respectively. For comparison, the $H_{c2}(t)$ in a Ba(Fe$_{0.93}$Co$_{0.07}$)$_2$As$_2$ single crystal (data taken from Nakajima *et al*. [19]) is also shown in figure 1(b) (open symbols). The values of $H_{c2}$ at *T* = 0 K estimated using the Werthamer-Helfand-Hohenberg formula, $H_{c2}(0) = -0.69\,T_c|dH_{c2}/dT|_{T=T_c}$, were ~360 and 650 kOe along the *c*- and *ab*-directions, respectively. These values are about 25% smaller than $H_{c2}(0)$ in single crystals of Co-doped BaFe$_2$As$_2$ [20]. From figure 1(b), the anisotropy factor $\gamma = H_{c2}^{ab}/H_{c2}^{c}$ of the film is ~2 and has weak temperature dependence, as shown in the inset of figure 1(b).

The Hall voltage was obtained by subtracting the transverse voltages measured with positive and negative magnetic fields. Figure 2(a) shows the Hall resistivity ($\rho_{xy}$) at several temperatures as a function of field in the Co-doped BaFe$_2$As$_2$ thin film. At all temperatures, the low-field $\rho_{xy}$ is positive and has a positive slope till around 2 kOe; subsequently $\rho_{xy}$ drops with field and becomes negative at high fields. With increasing temperature, the sign reversal of $\rho_{xy}$ sets in at progressively higher fields – varying from ~3 kOe at 25 K to ~16 kOe at 300 K. In our film, while the low-temperature $\rho_{xy}$ is linear in *H* (except at very low fields, viz., below 2 kOe), at high temperatures the *H*-linear part appears only at higher fields. By contrast, $\rho_{xy}$ is negative at all temperatures in single crystals of Co-doped BaFe$_2$As$_2$ and varies linearly with *H* [18,19]. The $\rho_{xy}$ variation in the film suggests a contribution arising from anomalous Hall effect (AHE) due to ferromagnetic Fe and/or Co. The Hall coefficient in a ferromagnet is empirically expressed as $\rho_{xy} = R_H H + 4\pi R_S M$, where $R_H$ and $R_S$ are the ordinary and anomalous Hall coefficients, respectively. In pure Fe and Co-doped Fe, it has been shown that the $H$-linear term is suppressed at lower fields by AHE [21]. Hence, we believe, our data shows a substantial influence of AHE in the transverse resistivity. We note that recently AHE was also reported in $FeSe_{1-x}Te_x$ films [22]. Shown in figure 2(b) (filled symbols) is the temperature dependence of the Hall coefficient $R_H$ obtained from the $H$-linear portion of $\rho_{xy}$. The sign of $R_H$ is negative in the whole temperature range, similar to single crystals [18,19,23] which indicates that the majority charge carriers are electrons. $R_H$ decreases gradually with decreasing temperature from 300 K to around 75 K beyond which $R_H$ has relatively weak temperature dependence. The value of $R_H$ at 300 K is three times larger than that just above $T_c$. The values of $R_H$ agree quite well with that in slightly overdoped single crystals [23].

Figure 3(a) shows the zero-field $I$-$V$ characteristics at different temperatures ranging from 2 K to 17 K. Measurable voltage appears once the driving force exceeds the pinning force in the superconductor. The critical current density $J_c$ was obtained from the $I$-$V$ plots using a voltage criterion of 1µV. The temperature dependence of $J_c$ is shown in figure 3(b). The films exhibit high $J_c$ values reaching up to 2.2 MA/cm$^2$ at 2 K. Figure 3(b) inset shows the magnetization hysteresis loop measured with $H // c$ at 3.5 K on a 1.4×1.55 mm$^2$ film using a vibration sample magnetometer. According to the Bean critical state model, $J_c$ (which is assumed to be field-independent) is given by

$$J_c = 20 \frac{\Delta M}{a\left(1 - \frac{a}{3b}\right)}$$

where $\Delta M$ is the difference between the magnetizations on the field-decreasing and field-increasing legs of the hysteresis loops and $a$ and $b$ are the film widths with $a < b$. Also shown in figure 3(b) main panel (star) is the self-field $J_c$ extracted from the hysteresis loop at 3.5 K. The extracted $J_c$ (2.5 MA/cm$^2$) is larger than that obtained from transport measurement.

Magneto-optical (MO) images were captured using the local-field dependent Faraday rotation of linearly polarized light in an in-plane magnetized garnet film placed above the sample. The intensity values in an MO image are directly related to the local axial magnetic induction $B_z$ [24]. MO images in the remanent state were obtained on a 0.5×0.5 mm$^2$ patterned film (figure 4(a)) after applying a field of 800 Oe which was subsequently decreased down to zero. Shown in figures 4(b)-(e) are MO images of the remanent state captured at several different temperatures. A defective portion across the left edge of the film (see figure 4(a)) causes a severe suppression of contrast in the MO image on this edge. The two lines, running from the top to the left and right, across the film in the MO images are scratches on the garnet indicator film. In these figures, the bright regions correspond to the trapped flux in the film. At all temperatures, the flux distribution is inhomogeneous which arises from extended defects in the film. These flux distribution images have no relation to dendritic flux patterns observed in other superconductors [25]. The trapped flux has maximum near the centre and along one of the diagonals of the film. The deviation from the ideal flux profiles, as expected from Bean's critical state model, is due to the presence of extended defects in the film. With increasing $T$, decreased pinning caused by thermal fluctuations gradually smears out the flux profile. In figure 4(f) we show the magnetic flux profiles ($B_z(x)$) at different temperatures in the thin film. The flux profile at each $T$ is obtained from the corresponding MO image taken along the dashed line represented in figure 4(b). In figure 4(f) the feature at around -150 µm and 300 µm in all curves are due to defects on the indicator film. The presence of extended defects gives rise to the irregular profiles. However, the overall flux profile is as expected for a thin film, with a sign change of the magnetic induction close to the edge of the film (Across this line, the actual film width is reduced due to the defective portion on the left edge). Within the Bean's critical state model for a thin film [26], $B_z(x)$ is related to $J_c$ as

$$B_z(x) = \frac{J_c 2d}{c} \ln\left\{\frac{[z^2 + (x-a)^2][z^2 + (x+a)^2]}{[z^2 + x^2]^2}\right\}$$

where $z$ is the height of the garnet film from the surface of the Co-doped $BaFe_2As_2$ film of width $2a$ and thickness $2d$. Using $z = 2$ μm for an optimal setting [27], a fit to $B_z(x)$ at 5 K using the above equation is shown in figure 4(f) (dashed line). The flux profiles in our sample do not match very well with the fit due to the asymmetry of the profile caused by extended defects in the film. The fit gives a value of $J_c = 1.3$ MA/cm$^2$. From the MO images, $J_c$ can also be estimated assuming $J_c \sim \Delta B_z/2d$ where $\Delta B_z$ is the difference in local induction between the center and edge of the film and $2d$ is the film thickness. At $T = 5$ K, this gives a value of $J_c \sim 3$ MA/cm$^2$. In comparison, at 5 K the $J_c$ obtained from transport measurements (figure 3(b)) is $\sim 2$ MA/cm$^2$, which is midway between the $J_c$ obtained from the flux profiles in two different ways. Thus, we can safely assume that the local $J_c$ is similar to that obtained by global transport measurements.

To summarize, resistivity, Hall coefficient, current-voltage characteristics, and magneto-optical imaging measurements were performed in epitaxial Co-doped $BaFe_2As_2$ thin films. The Hall resistivity of the film has a substantial contribution arising from anomalous Hall effect of ferromagnetic elements. Hence the $H$-linear part of the Hall resistivity differs at low and high temperatures. From transport measurements, the critical current density $J_c$ of the film is $\sim 2.2$ MA/cm$^2$ (at 2 K). The differential magneto-optical images of the remanent state also give values of $J_c$ similar to that obtained by transport measurement. However, the presence of extended defects is also demonstrated in the magneto-optical images. Elimination of such defects is vital for future improvement of film quality.

The woks at the Frontier Research Center were supported by the Japan Society for the Promotion of Science (JSPS), Japan, through "Funding Program for World-Leading Innovative R&D on Science and Technology (FIRST) Program".

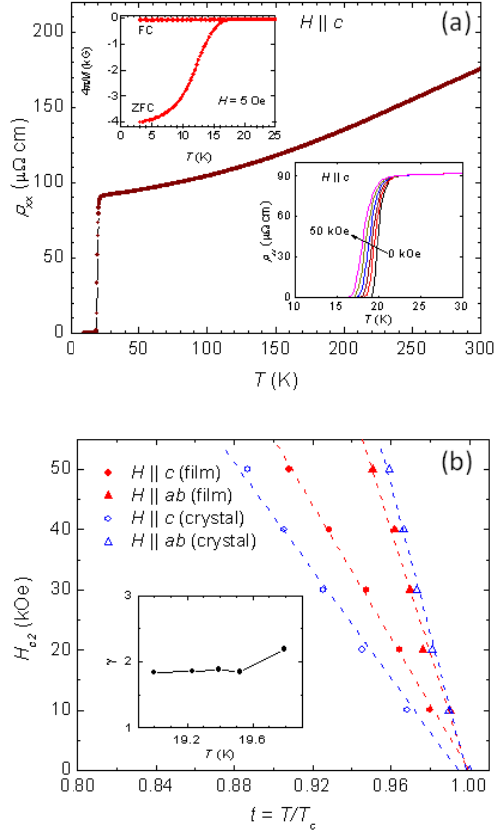

**Figure 1.** (a) The temperature dependence of the in-plane resistivity $\rho_{xx}$ in the Co-doped $BaFe_2As_2$ thin film at zero-field. The upper inset shows the magnetization of the film measured with $H$ = 5 Oe along the $c$-axis in ZFC and FC modes of measurement. The lower inset shows the variation of $\rho_{xx}$ near $T_c$ with magnetic field along the $c$-axis of the film for $H$ = 0, 10, 20, 30, 40, and 50 kOe. (b) Plot of the the upper critical field $H_{c2}$ along the $c$- and $ab$-directions versus temperature determined by the midpoint of resistive transition from the lower inset of (a) in Co-doped $BaFe_2As_2$ thin film (filled symbols) and in a single crystal (open symbols). Inset shows the temperature dependence of the anisotropy factor in the thin film.

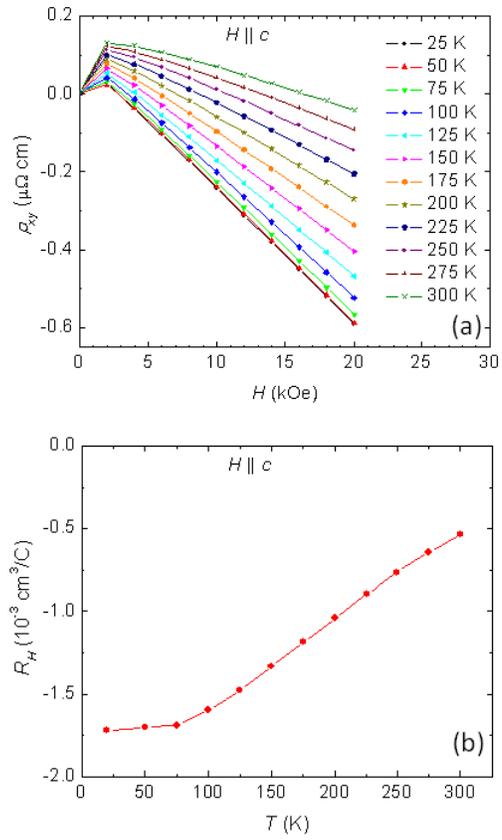

**Figure 2.** (a) The Hall resistivity $\rho_{xy}$ at several temperatures as a function of magnetic field. (b) The plot of the Hall coefficient $R_H$ with temperature obtained from (a).

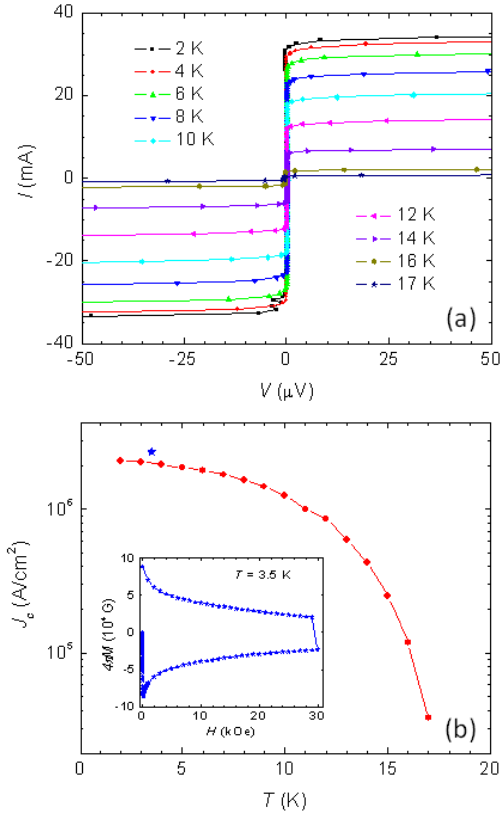

**Figure 3.** (a) The zero-field current-voltage characteristics in Co-doped BaFe$_2$As$_2$ thin film at $T$ = 2, 4, 6, 8, 10, 12, 14, 16, and 17 K. (b) The critical current density $J_c$ at different $T$ obtained from (a). The data point at 3.5 K (star) is obtained from the magnetization hysteresis loop shown in the inset.

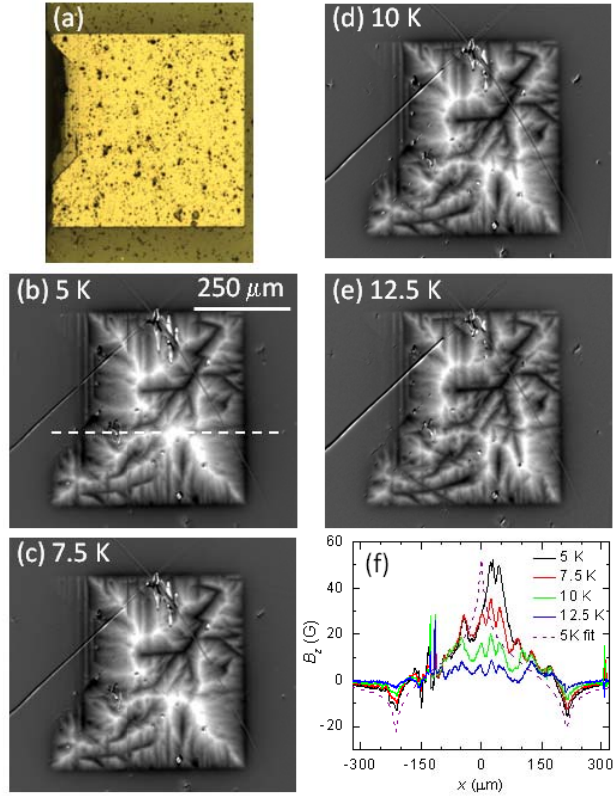

**Figure 4.** (a) Optical image of the film used for magneto-optical imaging. The differential magneto-optical images of the remanent state in Co-doped BaFe$_2$As$_2$ thin film at (b) 5 K, (c) 7.5 K, (d) 10 K, and (e) 12.5 K. (f) The local magnetic induction profiles at different $T$ taken along the dashed line illustrated in (b). The dashed line in (f) is a fit to the profile at 5 K using the critical state model.